\documentclass[12pt]{article}
\usepackage{color}
\usepackage[dvips]{graphicx}
\usepackage[latin1]{inputenc}
\usepackage{amsmath}
\usepackage{amssymb}
\newcommand{\be}{\begin{equation}}
\newcommand{\ee}{\end{equation}}
\newcommand{\beann}{\begin{eqnarray*}}
\newcommand{\eeann}{\end{eqnarray*}}
\newcommand{\bea}{\begin{eqnarray}}
\newcommand{\eea}{\end{eqnarray}}

\newcommand{\bdm}{\begin{displaymath}}
\newcommand{\edm}{\end{displaymath}}

\begin{document}

\centerline{\bf DILATON QUANTUM COSMOLOGY WITH}
\vspace{0.2cm}
\centerline{\bf A SCHR\"ODINGER-LIKE EQUATION}
\vspace{.70cm}
\centerline{J.C. Fabris\footnote{e-mail: fabris@pq.cnpq.br}$ ^{,a}$, F.T. Falciano\footnote{e-mail: ftovar@cbpf.br}$ ^{,b}$, J. Marto\footnote{e-mail: jmarto@ubi.pt}$ ^{,c}$,
N. Pinto-Neto\footnote{e-mail: nelsonpn@cbpf.br}$ ^{,b}$, P. Vargas Moniz\footnote{e-mail: pmoniz@ubi.pt}$ ^{,c}$}
\vspace{0.7cm}
\noindent
{a) Departamento de F\'{\i}sica, Universidade Federal do Esp\'{\i}rito Santo, ES, Brasil.\\
\\
b) ICRA, Centro Brasileiro de Pesquisas F\'{\i}sicas, RJ, Brasil.\\
\\
c) Departamento de F\'{\i}sica, Universidade da Beira Interior, Covilh\~a, Portugal.}
\vspace{0.5cm}
\date{\today}
\vspace{0.5cm}

\centerline{\bf Abstract}
A quantum cosmological model with radiation and a dilaton scalar field is analysed. The Wheeler-deWitt equation in the mini-superspace induces a Schr\"odinger equation, which can be solved. An explicit wavepacket is constructed
for a particular choice of the ordering factor. A consistent solution is possible only when the scalar field is a phantom field. Moreover, although the wavepacket is time dependent, a Bohmian
analysis allows to extract a bouncing behaviour for the scale factor.

%%%%%%%%%%%%%%%%%%%%%%%%%%%%%%%%%
%%%%%%%%%%%%%%%%%%%%%%%%%%%%%%%%%
%%%%%%%%%%%%%%%%%%%%%%%%%%%%%%%%%
%%%%%%%%%%%%%%%%%%%%%%%%%%%%%%%%%
%%%%%%%%%%%%%%%%%%%%%%%%%%%%%%%%%
\section{Introduction}

A scalar field is the simplest fundamental matter field that we may introduce in a cosmological model, since it adds essentially just one degree of freedom and it is naturally invariant under coordinate transformation. Even though, we can quote two different couplings of a scalar field: the minimal
coupling, leading to the so-called Einstein's frame; the non-minimal coupling, corresponding to the Jordan's frame. Of course, the minimal coupling is the most simple one. But, the non-minimal coupling
is one of the predictions of string theory: the scalar field emerging from this theory, called dilaton field, appears non-minimally coupled to the gravity term
\cite{copeland}. Moreover, being an interesting
alternative to the General Relativity theory, the Brans-Dicke theory \cite{bd}, is based on the non-minimal coupling.
\par
Both frames, the Einstein and Jordan ones, may be connected by a conformal transformation. There is an intensive discussion in the literature about the meaning of such connection, in the sense
that it may be just a mathematical mapping of one system into another, or may hide a deep physical meaning, see for example \cite{frame} and references therein. For some observables, it is clear that the predictions in one frame are not equivalent to
the predictions in the other frame. In quantum cosmology this problem has been treated in references \cite{nelson1,nelson2}, showing that it is possible - even at quantum level - to mapp the equations
in one frame into another frame, but with different predictions to the evolution of the universe.
\par
In references \cite{nelson1,nelson2} a system consisted of gravity and a scalar field, minimally or non-minimally coupled has been considered. In this case, the analysis present a major
challenge, typical of quantum cosmology: how to obtain predictions for the evolution of the universe as function of time? In fact, it is well known that, considering the Einstein-Hilbert action, in quantizing it
the resulting equation, the Wheeler-de Witt equation, has no time parameter due to the invariance of the gravitational system under time reparametrization.
The introduction of a scalar field does not change this situation. In those references, the $WKB$ method has been used in order to obtain predictions for the evolution of the universe from the wavefunction
determined from the (timeless) Wheeler-de Witt in the minisuperspace. Concerning this question, see reference \cite{nelsonr} and references therein.
\par
Another possibility to obtain the time evolution of a quantum cosmological system in the minisuperspace is to introduce matter in the form of a fluid, employing the Schutz description
based in terms of potential that conveys the degrees of freedom of the fluid \cite{schutz1,schutz2}.
It has been shown that such description leads, always in the minisuperspace, to a Schr\"odinger-like equation, with the degrees of freedom of the matter playing the r\^ole of time.
This proposal has been presented in reference \cite{rubakov}, and extensively analysed in reference \cite{lemos}.
\par
Our aim in this work is to address the problem of time evolution of the universe in the presence of a dilaton-like field from a quantum cosmological perspective. Since we are interested mainly in the primordial universe, a radiative fluid
will be introduced. This may allow to recover the time variable through the employment of the Schutz formulation. Performing a conformal transformation (which does not affect the radiative fluid,
 since it is conformal invariant), we re-write the dilaton-gravity system in the Einstein's frame, from which the Wheeler-de Witt equation in the minisuperspace is constructed, resulting
in a Schr\"odinger-like equation. From this Schr\"odinger-like equation, an explicit solution
is obtained by using a specific ordering factor. A wavepacket is determined, but its norm is time-dependent. Hence, it does not fit in the usual
 many-worlds interpretation of quantum mechanics \cite{tipler,nelsonr}, but it admits a Bohmian analysis \cite{nelsonr,holland}. The Bohmian trajectories contain universes that are singurality-free.
\par
This paper is organized as follows. In the next section, we quantize the dilaton-gravity-radiation system in the mini-superspace. In section 3, a wavepacket is constructed and
a (formal) many-world analysis is performed. In section 4 the Bohmian trajectories are studied. In section 5 we present our conclusions.

\section{Dilaton-gravity system with radiative fluid}

Let us consider the non-minimal coupling of gravity and a scalar field, represented by the Brans-Dicke theory:

\begin{eqnarray}
\label{classLD}
{\cal L} = \sqrt{-\tilde g}\phi\biggr\{\tilde R -
\tilde \omega\frac{\phi_{;\rho}\phi^{;\rho}}{\phi^2}\biggl\} + {\cal L}_m,
\end{eqnarray}
where $L_m$ is the matter Lagragian supposed to be conformal invariant (a radiative fluid).
In the strict string dilatonic case we have $\tilde\omega = - 1$. This defines the theory in the Jordan's frame.
Performing a conformal transformation such that $g_{\mu\nu} = \phi^{-1}\tilde g_{\mu\nu}$, we transpose the action (\ref{classLD}) to the corresponding expression written in the
Einstein's frame:
\begin{eqnarray}\label{class_L}
{\cal L} = \sqrt{-g}\biggr\{R -
\omega\frac{\phi_{;\rho}\phi^{;\rho}}{\phi^2}\biggl\} + {\cal L}_m,
\end{eqnarray}
where $\omega = \tilde\omega + 3/2$.

Restricting ourselves to the FLRW metric, in an appropriate coordinate system, the line element can be written as
\begin{equation}\label{FLRW}
ds^2 = N(t)^2dt^2 - a(t)^2 \gamma_{ij}dx^i dx^j
\end{equation}
where $N(t)$ is the lapse function, $a(t)$ is the scale factor and $\gamma_{ij}$ is the induced metric of the homogeneous and isotropic spatial hypersurfaces with curvature $k=0,\pm1$.
From now on, we will fix $k = 0$.
With this metric, the gravitational Lagrangian becomes,
\begin{eqnarray}
{\cal L}_G = \frac{V_0a^3}{N}\biggr\{-6\biggr[\frac{\ddot a}{a} + \biggr(\frac{\dot a}{a}\biggl)^2 - \frac{\dot a}{a}\frac{\dot N}{N}\biggl] - \omega
\frac{\dot \phi^2}{\phi^2}\biggl\}\quad ,
\end{eqnarray}
where $V_0$ is a constant and can be interpreted as the physical volume of the compact universe divided by $a^3$. Since we shall have an identical multiplicative constant in front of the matter lagrangian we can drop it from our analysis (this can also be understood as a normalization of the fields). Hence, discarding a surface term, the gravitational Lagrangian can be written as,
\begin{eqnarray}
{\cal L}_G = \frac{1}{N}\biggr\{6a\dot a^2 - \omega a^3\frac{\dot \phi^2}{\phi^2} \biggl\}.
\end{eqnarray}
Defining,
\begin{equation}
\sigma = \sqrt{|\omega|}\ln\phi,
\end{equation}
we obtain,
\begin{eqnarray}
{\cal L}_G = \frac{1}{N}\biggr\{6a\dot a^2 - \epsilon a^3\dot\sigma^2 \biggl\},
\end{eqnarray}
where $\epsilon = \pm 1$ according $\omega$ is positive (upper sign) or negative (lower sign).
The canonical momenta associated with the scale factor and the scalar field are respectively:
\begin{eqnarray}
\label{momentum}
p_a = 12\frac{a\dot{a}}{N}\quad &,& \quad p_\sigma = - 2\epsilon \frac{a^3\dot\sigma}{N} \qquad.
\end{eqnarray}
This leads to the following expression in terms of the conjugate momentum:
\begin{eqnarray}
{\cal L}_G = p_a\dot{a}+p_\sigma\dot{\sigma}-N\biggr\{\frac{1}{24}\frac{p_a^2}{a} - \epsilon
\frac{p_\sigma^2}{4a^3}\biggl\}\qquad.
\end{eqnarray}
Considering a radiative matter component (for the computation of the conjugate momentum associated with the fluid, see references \cite{rubakov,lemos}), the total Hamiltonian is:
\begin{eqnarray}
H = N\biggr\{\frac{1}{24}\frac{p_a^2}{a} -
\epsilon\frac{p_\sigma^2}{4a^3}-
\frac{p_T}{a}\biggl\}.
\end{eqnarray}
The resulting Schr\"odinger equation is,
\begin{equation}
\label{se}
- \frac{\partial^2\Psi}{\partial a^2} + \frac{\epsilon}{a^2}\frac{\partial^2\Psi}{\partial\sigma^2} = i \frac{\partial\Psi}{\partial T},
\end{equation}
where we made the redefinition $\frac{\sigma}{\sqrt{6}} \rightarrow \sigma$ and $\frac{T}{24} \rightarrow T$.
\par
Two questions related to (\ref{se}) appear immediately. First, what is the sign of $\epsilon$? If it is positive, we have a hyperbolic Schr\"odinger equation. This implies that the "energy" $E$ is not positive defined, as it can be seen by multiplying
the Schr\"odinger equation by $\Psi^*$ and integrating in $\sigma$ and $a$. Moreover, in this case, the argument of the Bessel function can become imaginary, what may not be very serious, but may
pose some problems to construct the wavepacket, since the integration in $E$ must be done along all real axis, and for $E < 0$ the Bessel function becomes modified Bessel function.
On the other hand, if the sign is $-1$ the positivity of $E$ seems to be assured, and the construction of the wavepacket seems not problematic, since the integration on $E$ is done only in the
positive semi-axis. For all these reasons, for the moment, we consider the case $\epsilon = - 1$. It corresponds to a phantom scalar behaviour induced to the dilaton field.
\par
Second, we have also not taken into account a possible factor ordering. If such factor ordering is introduced, with $\epsilon = - 1$, we have the following Schr\"odinger equation:
\begin{equation}\label{WDW_factor_p}
\frac{\partial^2\Psi}{\partial a^2} + \frac{p}{a}\frac{\partial \Psi}{\partial a} + \frac{1}{a^2}\frac{\partial^2\Psi}{\partial\sigma^2} = - i \frac{\partial\Psi}{\partial T}\quad.
\end{equation}

Even though there is no unique way to chose the factor ordering, a possible consistent and adequate choice is the covariant ordering which is invariant through fields re-definitions. The minisuperspace for our model is two dimensional. Thus, using the covariant factor ordering, we can re-write the Wheeler-DeWitt equation in a Klein-Gordon like equation by defining the minisuperspace metric such that eq.~(\ref{WDW_factor_p}) reads
\begin{equation}\label{WDW_cov}
\frac{1}{\sqrt{g}}\partial_{\mu}\left(
\sqrt{g}g^{\mu \nu}\partial_{\nu}
\right)\Psi
= - i \frac{\partial\Psi}{\partial T}\quad,
\end{equation}
where $\mu$, $\nu=0,1$ and the coordinates represents the fields, i.e. $x^0=a$ and $x^1=\sigma$. The only possible value for $p$ in eq.~(\ref{WDW_factor_p}) which allows the covariant factor ordering is $p=1$. In this case, we have
\begin{eqnarray}
\label{felipe}
g_{\mu \nu}=\left(
\begin{array}{cc}
1&0\\
0&a^2
\end{array}
\right)\quad \Rightarrow\qquad \sqrt{g}=a
\end{eqnarray}
and the Wheeler-DeWitt equation reads
\begin{equation}\label{WDW_p1}
\left[\frac{1}{a}\frac{\partial}{\partial a}\left(a \frac{\partial}{\partial a}\right) + \frac{1}{a^2}\frac{\partial^2}{\partial\sigma^2}\right] \Psi= - i \frac{\partial\Psi}{\partial T}\quad.
\end{equation}

This equation can be solved using the technique of separation of variables. Hence, through the ansatz
\begin{equation}
\Psi(a,\sigma,T) = \phi(a)e^{ik\sigma}e^{- iET},
\end{equation}
we obtain,
\begin{equation}
\phi'' + \frac{\phi'}{a} + \biggr\{E - \frac{k^2}{a^2}\biggl\}\phi = 0,
\end{equation}
where the prime means derivative with respect to $a$. Noticing that this is just a Bessel's equation, its solution reads
\begin{equation}
\label{sol}
\Psi = AJ_\nu(\sqrt{E}a)e^{ik\sigma}e^{-iET}, \quad \nu = k\quad ,
\end{equation}
with $A$ being a normalization constant.

%%%%%%%%%%%%%%%%%%%%%%%%%%%%%%%%%
%%%%%%%%%%%%%%%%%%%%%%%%%%%%%%%%%
%%%%%%%%%%%%%%%%%%%%%%%%%%%%%%%%%
%%%%%%%%%%%%%%%%%%%%%%%%%%%%%%%%%
%%%%%%%%%%%%%%%%%%%%%%%%%%%%%%%%%
\section{Wavepacket and expectation values}

Let us construct a wavepacket by choosing conveniently the function $A = A(E,k)$.
We have, for (\ref{sol}),
\begin{equation}
\Psi(a,\sigma,T) = \int_0^\infty\int_0^\infty r^{\nu + 1}e^{-(\gamma + iT)r^2}e^{- (\alpha - i\sigma)k}J_k(ra)dk\,dr,
\end{equation}
where $r = \sqrt{E}$ and $\alpha,\gamma$ are postive constants.
Using the formula (6.631-4) from \cite{grad}, the final result is:
\begin{equation}
\Psi(a,\sigma,T) = C\frac{e^{- \frac{a^2}{4B(T)}}}{B(T)\,g_\alpha(a,B,\sigma)},
\end{equation}
where
\begin{equation}
B(T) = (\gamma + i\,T), \quad g_\alpha(a,\sigma,T) = - \alpha + \,\ln\biggr[\frac{a}{2B(T)}\biggl] \pm i\sigma,
\end{equation}
and $C$ is a normalization constant. Remark that, in order to give physical meaning for this wave packet, the condition $\Re\, g_\alpha(a,\sigma,T) < 0$ must be satisfied, assuring that
the integral in the separation parameter $k$ is convergent.
\par
Let us calculate this normalization constant. Unitarity requires,
\begin{eqnarray}
N = \int \Psi^*\Psi da\,d\sigma = 1.
\end{eqnarray}
But,
\begin{eqnarray}
\Psi^*\Psi = C^2 \frac{e^\frac{-\gamma a^2}{2B^*B}}{2B^*B}\frac{1}{g^*g}.
\end{eqnarray}
Remark that,
\begin{eqnarray}
B = \gamma + iT = De^{i\theta},
\end{eqnarray}
where
\begin{eqnarray}
D &=& \sqrt{\gamma^2 + T^2} = \sqrt{B^*B},\\
\label{s}
\theta &=& \arctan\biggr(\frac{T}{\gamma}\biggl).
\end{eqnarray}
Hence,
\begin{equation}
g^*g = h^2 + (\sigma - \theta)^2, \quad h = - \alpha + \ln\biggr(\frac{a}{2(B^*B)^\frac{1}{2}}\biggl).
\end{equation}
For the scalar field $\sigma$ and for $a$ we must compute a double integral. In doing so, we must fix the range of $\sigma$. In principle, it is
from $- \infty < \sigma < + \infty$, implying $0 \leq \phi < + \infty$, that is, a positive gravitational coupling.
The norm of the wave function becomes
\begin{eqnarray}
N = \frac{C^2}{(2B\,B^*)^{1/2}}\int_0^\infty\int_{-\infty}^{+\infty}\frac{e^{-\gamma\frac{a^2}{2BB^*}}}{(2B^*B)^\frac{1}{2}}\frac{d\sigma}{[h^2 + (\sigma - \theta)^2]}da.
\end{eqnarray}
Under the substitutions,
\begin{eqnarray}
u = \frac{a}{(2B^*B)^\frac{1}{2}},\quad v = - \theta + \sigma,
\end{eqnarray}
the integral above becomes,
\begin{equation}
N = \frac{C^2}{(B\,B^*)^{1/2}}\int_0^\infty\int_{-\infty}^{+\infty}\frac{e^{-\gamma u^2}}{h^2 + v^2}du\,dv.
\end{equation}
But,
\begin{equation}
\int_{-\infty}^{+\infty}\frac{dv}{h^2 + v^2} = \frac{\pi}{h}.
\end{equation}
Hence,
\begin{equation}
N =\frac{C^2}{(B\,B^*)^{1/2}} \pi\int_0^\infty \frac{e^{-\gamma u^2}}{\alpha + \ln\biggr(\frac{u}{2}\biggl)}du =\frac{C^2}{(B\,B^*)^{1/2}} \pi I_1 ,
\end{equation}
where $I_1$ is the definite integral,
\begin{equation}
I_1 = \int_0^\infty \frac{e^{-\gamma u^2}}{\alpha + \ln\biggr(\frac{u}{2}\biggl)}du.
\end{equation}
The wavepacket determined above is time dependent and does not convey unitarity.
\par
As a formal exercise, even in absence of unitarity we evaluate the expectation value of the scalar field and of the scale factor for this wavepacket:
\begin{eqnarray}
\label{escalar}
<\sigma>_T &=& \frac{\int_0^\infty\int_{-\infty}^{+\infty}\Psi^*\,\sigma\,\Psi\,da\,d\sigma}{\int_0^\infty\int_{-\infty}^{+\infty}\Psi^*\Psi\,da\,d\sigma},\\
<a>_T &=& \frac{\int_0^\infty\int_{-\infty}^{+\infty}\Psi^*\,a\,\Psi\,da\,d\sigma}{\int_0^\infty\int_{-\infty}^{+\infty}\Psi^*\Psi\,da\,d\sigma}.
\end{eqnarray}
Let us first evaluate the expectation value for the scalar field.
In the numerator of (\ref{escalar}), we have:
\begin{equation}
\int_{-\infty}^{+\infty}\frac{\sigma d\sigma}{h^2 + (\sigma - \theta)^2} = \int_{-\infty}^{+\infty}\frac{(v + \theta) dv}{h^2 + v^2}.
\end{equation}
The first integral is zero, and the second one leads to,
\begin{equation}
\int_{-\infty}^{+\infty}\frac{\sigma d\sigma}{h^2 + (\sigma - \theta)^2} =  \theta\int_{-\infty}^{+\infty}\frac{dv}{h^2 + v^2} = \theta\frac{\pi}{h}.
\end{equation}
Re-inserting in the expression for the scale factor, we find:
\begin{equation}
<\sigma>_T = \arctan\biggr(\frac{T}{\gamma}\biggl).
\end{equation}
For the scale factor, integrating in $\sigma$ we have:
\begin{equation}
<a>_T = \pi\int_0^\infty \frac{a}{h} \frac{e^{- 2\gamma\frac{a^2}{B^*B}}}{B^*B} da.
\end{equation}
Performing, as before, the substitution
\begin{equation}
\frac{a}{\sqrt{B^*B}} = \frac{a}{\sqrt{\gamma^2 + T^2}} = u,
\end{equation}
we obtain:
\begin{equation}
<a>_T = \sqrt{\gamma^2 + T^2}\int_0^\infty \frac{u}{h} e^{- 2\gamma u^2} du = \frac{I_2}{I_1}\sqrt{\gamma^2 + T^2},
\end{equation}
where
\begin{equation}
I_2 = \int_0^\infty \frac{u}{h} e^{- 2\gamma u^2} du.
\end{equation}
Hence,
\begin{eqnarray}
<a>_T &=& a_0(\gamma^2 + T^2)^{1/2},
\end{eqnarray}
which is formally the same solution as when there is no scalar field \cite{lemos}. Remark that the norm of the wavepacket is time dependent and a unitary framework can not be
invoked.
\par
The fact that the result for the scale factor is the same, essentially, as when the scalar field is absent may be seen as contradictory. But,
we remark that in the Friedmann equation it appears the logarithmic derivative of the scale factor and the derivative of the scalar field.
Hence, the presence of the scale field just change the value of $a_0$ with respect to the case it is absent.

%%%%%%%%%%%%%%%%%%%%%%%%%%%%%%%%%
%%%%%%%%%%%%%%%%%%%%%%%%%%%%%%%%%
%%%%%%%%%%%%%%%%%%%%%%%%%%%%%%%%%
%%%%%%%%%%%%%%%%%%%%%%%%%%%%%%%%%
%%%%%%%%%%%%%%%%%%%%%%%%%%%%%%%%%
\section{Bohmian trajectories}

In order to extract some physical insight on the model developed until here, given the nature of the wavepacket employed in the previous section, we will use a Bohmian analysis \cite{nelsonr}. Note that the computation of the Bohmian trajectories makes sense even in absence of unitarity, this being one of the main features of the ontological interpretation of quantum mechanics.

The Bohm-de Broglie interpretation is most easily understood through the polar form of the wave function. Indeed, decomposing the wave function as $\Psi=Re^{iS}$, the Wheeler-DeWitt equation (\ref{WDW_p1}) split in two real non-linear equations for the two real function $R(a,\sigma,T)$ and $S(a,\sigma,T)$
\begin{eqnarray}
&&\frac{\partial R^2}{\partial T}+\frac1a\frac{\partial }{\partial a}\left(R^2\ 2a\frac{\partial S}{\partial a}\right)+\frac1a\frac{\partial }{\partial \sigma}\left(R^2\frac{2}{a}\frac{\partial S}{\partial \sigma}\right)=0\label{eq_cont}\\
&&\frac{\partial S}{\partial T}+\left(\frac{\partial S}{\partial a}\right)^2+\left(\frac{1}{a}\frac{\partial S}{\partial \sigma}\right)^2+Q=0 \label{HJ_mod}
\end{eqnarray}
with
\begin{equation}
Q\equiv -\frac{1}{R}\left[\frac1a\frac{\partial}{\partial a}\left(a\frac{\partial R}{\partial a}\right)+\frac{1}{a^2}\frac{\partial^2 R}{\partial \sigma^2}\right]
\end{equation}

Equation (\ref{HJ_mod}) is a modified Hamilton-Jacobi equation with the presence of the quantum potential $Q$. Through this equation, we can identify the momenta as
\begin{eqnarray}\label{bohm_momenta}
P_a=\frac{\partial S}{\partial a}\qquad&,&\qquad P_\sigma=\frac{\partial S}{\partial \sigma}\quad .
\end{eqnarray}
The metric in the minisuperspace (\ref{felipe}) allows to define covariant derivatives, see reference \cite{felipe}.
Therefore, equation (\ref{eq_cont}) represents a continuity equation written in that minisuperspace, with $R^2$ playing the r\^ole of a probability density and we can directly read the associated velocity within each parenthesis, i.e.
\begin{displaymath}\label{guidance_rel}
v^\mu=g^{\mu \nu}\frac{\partial S}{\partial x^\nu}\quad \Rightarrow \qquad
 \dot{a}=2\frac{\partial S}{\partial a} \qquad , \qquad \dot{\sigma}=\frac{2}{a^2}\frac{\partial S}{\partial \sigma}
 %\left\{ \begin{array}{c} \dot{a}=\frac{\partial S}{\partial a} \dot{\sigma}=\frac{1}{a^2}\frac{\partial S}{\partial \sigma} \end{array}\right.
\end{displaymath}

The wavefunction can be written as,
\begin{eqnarray}
\Psi(a,\sigma,T) &=& A \frac{e^{-\frac{a^2}{4B}}}{g\,B} = A \frac{e^{-\frac{a^2B^*}{4BB^*}}}{g^*g\,B^*B}g^*\,B^*,\nonumber\\
&=& f(a,\sigma,T)e^{iS(a,\sigma,T)},
\end{eqnarray}
with,
\begin{eqnarray}
f(a,\sigma,T) &=& A \frac{e^{-\frac{a^2\gamma}{4BB^*}}}{\sqrt{g^*g\,B^*B}},\\
S(a,\sigma,T) &=& \frac{a^2 T}{4(\gamma^2 + T^2)} - \arctan\biggr(\frac{T}{\gamma}\biggl) + \arctan\biggr(\frac{- \theta + \sigma}{h}\biggl).
\end{eqnarray}
Remembering that,
\begin{eqnarray}
\label{h}
h = - \alpha + \ln\biggr(\frac{a}{2\sqrt{\gamma^2 + T^2}}\biggl) = \Re \,g_T(a,\sigma).
\end{eqnarray}
and that, for a radiative fluid $N = a$ in (\ref{momentum}) \cite{lemos}, taking into account the time and field redefinitions,
we have the following expressions for the Bohmian trajectories:
\begin{eqnarray}
\label{bt1}
\dot a &=& \frac{a T}{(\gamma^2 + T^2)} + \frac{2}{a}\frac{\theta - \sigma}{h^2 + (\theta - \sigma)^2},\\
\label{bt2}
\dot\sigma&=& \frac{2}{a^2} \frac{h}{h^2 + ( \theta- \sigma)^2},
\end{eqnarray}
where $\theta$ and $h$ are given by (\ref{s},\ref{h}).
These equations must be integrated numerically\footnote{It can already be stated that the solutions are not equivalent to the expectation values found before, a consequence
of the absence of unitarity: when the wavepacket is unitary, the Bohmian trajectories reproduces the expectation values.}.
\par
The numerical solutions of (\ref{bt1},\ref{bt2}) provide an interesting perspective depending on the initial conditions and on the free parameters of the model, namely $\gamma$ and $\alpha$.
We can show that there is at least a class of non-singular solutions, satisfying the wavepacket conditions of consistency.
More precisely, in figure 1, the a non-singular solution is displayed. The function $h$ is also plotted, showing that the condition for the convergence of the wavepacket is satisfied.
In the same figure we plot the results obtained use the expectation value of the previous section, and the numerical solutions for the Bohmian trajectories. The Bohmian trajectories
predict an asymmetric bounce, while the expectation value display a perfect symmetric bounce.

\begin{center}
\begin{figure}[!t]
\begin{minipage}[t]{0.3\linewidth}
\includegraphics[width=\linewidth]{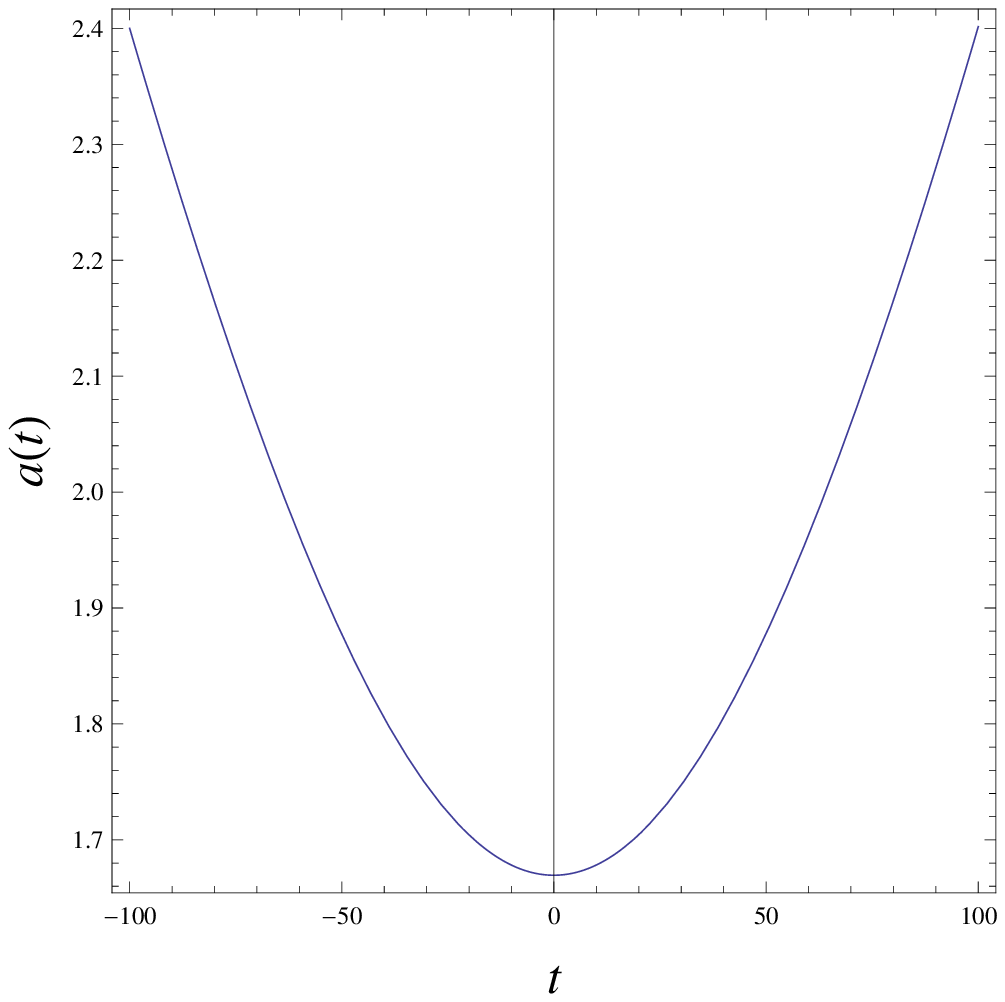}
\end{minipage} \hfill
\begin{minipage}[t]{0.312\linewidth}
\includegraphics[width=\linewidth]{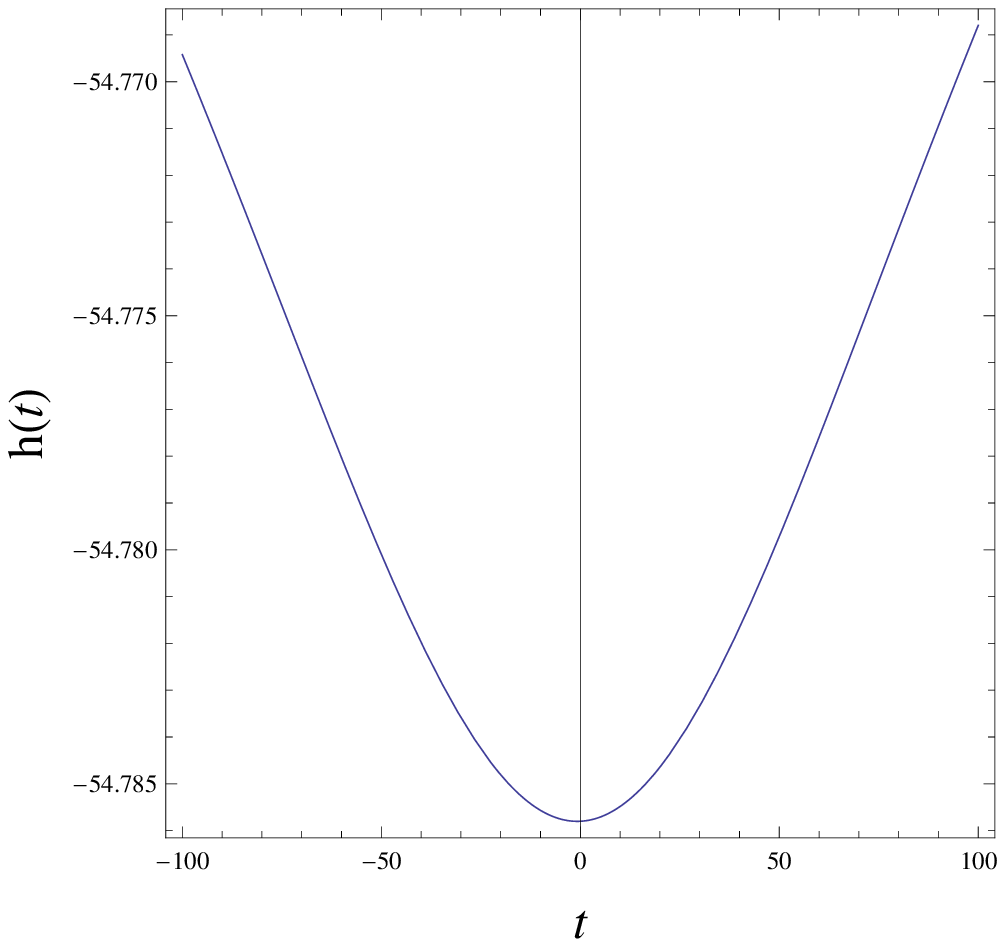}
\end{minipage} \hfill
\begin{minipage}[t]{0.3\linewidth}
\includegraphics[width=\linewidth]{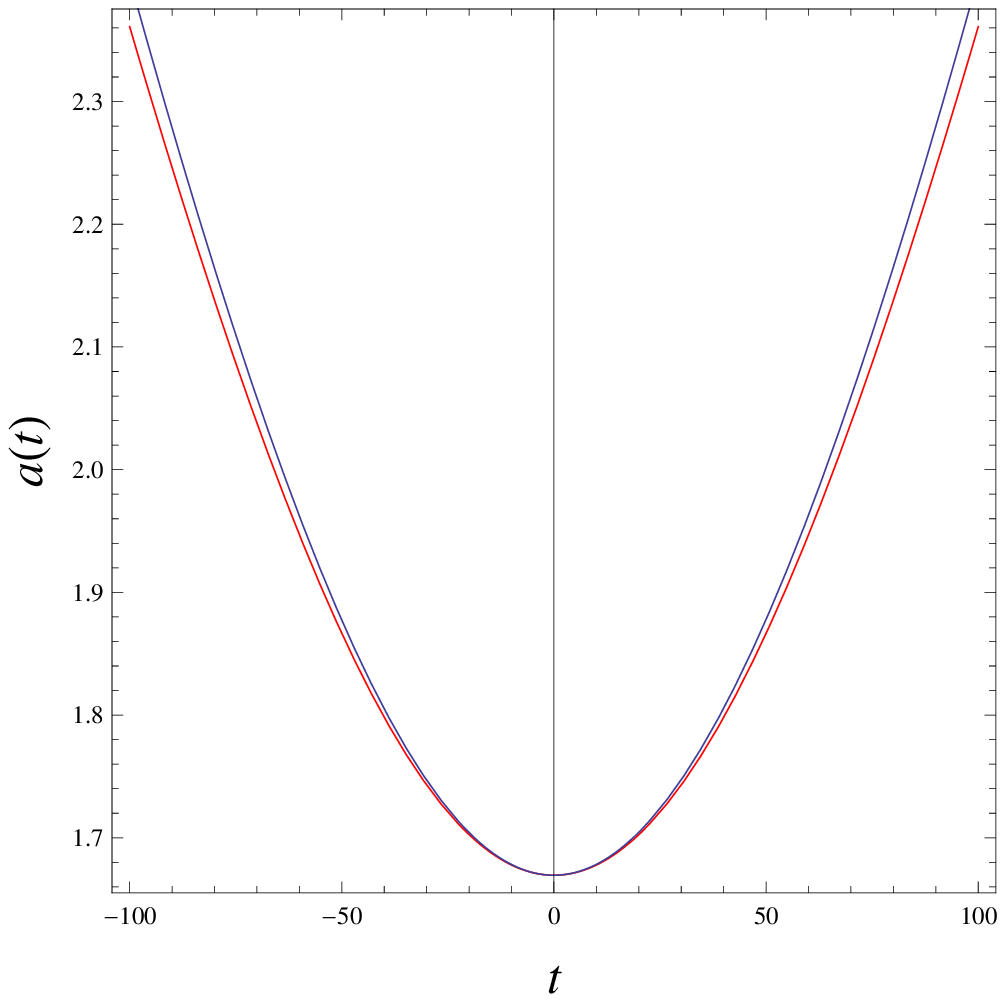}
\end{minipage} \hfill
\caption{In the left panel, the behaviour of the scale factor for $\gamma = 100$ and $\alpha = 50$ is shown. In the central panel, the function $h$ is displayed. In the
right panel, the solution obtained using the Bohmian trajectories (blue) is compared with the solution using the expectation value (red).}
\end{figure}
\end{center}

%%%%%%%%%%%%%%%%%%%%%%%%%%%%%%%%%
%%%%%%%%%%%%%%%%%%%%%%%%%%%%%%%%%
%%%%%%%%%%%%%%%%%%%%%%%%%%%%%%%%%
%%%%%%%%%%%%%%%%%%%%%%%%%%%%%%%%%
%%%%%%%%%%%%%%%%%%%%%%%%%%%%%%%%%
\section{Conclusions}

The possibility to make predictions and subsequently extract a family of solutions (trajectories in the minisuperspace) is a recurrent problem in quantum cosmology.
Although the well-known $WKB$ method, including decoherence features, has provided a vast range of particular results, there is an alternative framework by means of transforming
the $WdW$ equation into a Schr\"odinger equation, with the time variable induced by a matter component \cite{schutz1,schutz2,rubakov}. Quite interesting results have appeared
in the literature \cite{lemos} using this procedure.
\par
In this paper, we investigated how we could obtain some predictions with that alternative framework, using a system characterized by a dilaton field and radiation expressed using
the Schutz variables. The dynamics of the radiative fluid implies a time variable.
\par
We found that, in order the Schr\"odinger equation to be elliptic, leading to a positive energy spectrum, the dilaton field must have a phantom behaviour. On the other hand, the construction of a quasi gaussian superposition, conjugated with some convergence criteria, lead to a wavepacket with a time dependent norm and hence unitarity could not be invoked.
Nevertheless, this still allows us to investigate this wavepacket under a Bohmian perspective. Interestingly, we found a bouncing behaviour for the scale factor, i.e., the singularity is
avoided. In this sense we extend the results of reference \cite{brasil}, where the imposition of unitarity led to a conclusion that the dilaton field should be
constant. At same time, such approach open new perspective concerning anisotropic quantum models \cite{brasilbis}.
\par
There are still many open directions to explore in this program. In particular, we must look for, (i) other wavepackets, presumably by using numerical methods, (ii) study the dynamics in the Jordan frame.
We hope to address these problems in a future work.

{\bf Acknowledgements:} We thank CNPq for partial financial support. PVM is grateful to CENTRA-IST for financial assistance. He also wants to thank UFES, where this work was completed, for hospitability


\begin{thebibliography}{}
\bibitem{copeland} J.E. Lidsey, D. Wands and E. J. Copeland, Phys. Rep. {\bf 337}, 343(2000).
\bibitem{bd} C. Brans and R. H. Dicke, Phys. Rev. {\bf 124}, 925(1961).
\bibitem{frame} E. Alvarez and J. Conde, Mod. Phys. Lett. {\bf A17}, 413(2002).
\bibitem{nelson1} R. Colistete Jr, J.C. Fabris and N. Pinto-Neto, Phys. Rev. {\bf D57}, 4707(1998).
\bibitem{nelson2} J.C. Fabris, N. Pinto-Neto and A. Velasco, Class. Quan. Grav. {\bf 16}, 3807(1999).
\bibitem{nelsonr} N. Pinto-Neto, {\it Quantum cosmology}, in {\bf Cosmology and Gravitation}, edited by M. Novello, \'Editions Fronti\`eres, Gif-sur-Yvette(1996).
\bibitem{tipler} F. Tipler, Phys. Rep. {\bf 137}, 231(1986).
\bibitem{holland} P.R. Holland, {\bf The Quantum Theory of Motion}, Cambridge University Press, Cambridge(1983).
\bibitem{schutz1} B. F. Schutz, Phys. Rev.{\bf D2}, 2762(1970).
\bibitem{schutz2} B. F. Schutz, Phys. Rev. {\bf D4}, 3559(1971).
\bibitem{rubakov} V. G. Lapchinskii and V. A. Rubakov, Theor. Math. Phys. {\bf 33}, 1076(1977).
\bibitem{lemos} F.G. Alvarenga, J.C. Fabris, N.A. Lemos and G.A. Monerat, Gen. Rel. Grav. {\bf 34}, 651(2002).
\bibitem{grad} I.S. Gradshteyn and I.M. Ryzhik, {\bf Table of Integrals, Series, and Products}, Academic Press, San Diego(2007).
\bibitem{felipe} F.T. Falciano and  N. Pinto-Neto, Phys. Rev. {\bf D79}, 023507(2009).
\bibitem{brasil} F.G. Alvarenga, A.B. Batist and, J.C. Fabris, Int. J. Mod. Phys. {\bf D14}, 291(2005).
\bibitem{brasilbis} F.G. Alvarenga, A.B. Batista, J.C. Fabris and S.V.B. Gon\c{c}alves, Gen. Rel. Grav. {\bf 35}, 1659(2003).
\end{thebibliography}
\end{document}